\documentclass[prx, 10pt, twocolumn, floatfix, superscriptaddress]{revtex4-1}
\usepackage{amsmath} 
\usepackage{graphicx} 
\usepackage{verbatim} 
\usepackage{color} 
\usepackage{subfigure} 
\usepackage{hyperref} 
\usepackage{sidecap}
\usepackage{tikz}
\usepackage{amsmath,bm}
\usepackage{amsfonts}
\usepackage{mathtools}
\usepackage{epstopdf}
\usepackage{float}

\begin{document}

\title{Experimental realization of a Dirac monopole through the decay of an isolated monopole}
\author{T. Ollikainen}
\email{tuomas.ollikainen@aalto.fi}
\affiliation{QCD Labs, COMP Centre of Excellence, Department of Applied Physics, Aalto University, P.O. Box 13500, FI-00076 Aalto, Finland}
\affiliation{Department of Physics and Astronomy, Amherst College, Amherst, MA 01002-5000, USA}
\author{K. Tiurev}
\affiliation{QCD Labs, COMP Centre of Excellence, Department of Applied Physics, Aalto University, P.O. Box 13500, FI-00076 Aalto, Finland}
\author{A. Blinova}
\affiliation{Department of Physics and Astronomy, Amherst College, Amherst, MA 01002-5000, USA}
\author{W. Lee}
\altaffiliation{Present address: Department of Physics, Princeton University, Princeton, NJ 08544, USA}
\affiliation{Department of Physics and Astronomy, Amherst College, Amherst, MA 01002-5000, USA}
\author{D. S. Hall}
\affiliation{Department of Physics and Astronomy, Amherst College, Amherst, MA 01002-5000, USA}
\author{M. M\"ott\"onen}
\affiliation{QCD Labs, COMP Centre of Excellence, Department of Applied Physics, Aalto University, P.O. Box 13500, FI-00076 Aalto, Finland}
\affiliation{University of Jyv\"askyl\"a, Department of Mathematical Information Technology, P.O. Box 35, FI-40014 University of Jyv\"askyl\"a, Finland}


\begin{abstract}
We experimentally observe the decay dynamics of deterministically created isolated monopoles in spin-1 Bose--Einstein condensates. As the condensate undergoes a change between magnetic phases, the isolated monopole gradually evolves into a spin configuration hosting a Dirac monopole in its synthetic magnetic field. We characterize in detail the Dirac monopole by measuring the particle densities of the spin states projected along different quantization axes. Importantly, we observe the spontaneous emergence of nodal lines in the condensate density that accompany the Dirac monopole. We also demonstrate that the monopole decay accelerates in weaker magnetic field gradients.
\end{abstract}

\maketitle

\section{\label{sec:introduction}Introduction}
Spinor Bose--Einstein condensates (BECs) of alkali-atom gases~\cite{Anderson:1995,Davis:1995} offer an exceptionally versatile platform to study various topological defects~\cite{Ueda:2014}. These defects are qualitatively nontrivial configurations in the order parameter fields of BECs and, by definition, they are robust against perturbations. 

\begin{figure}[t!]
\centering
\includegraphics[width=0.475\textwidth]{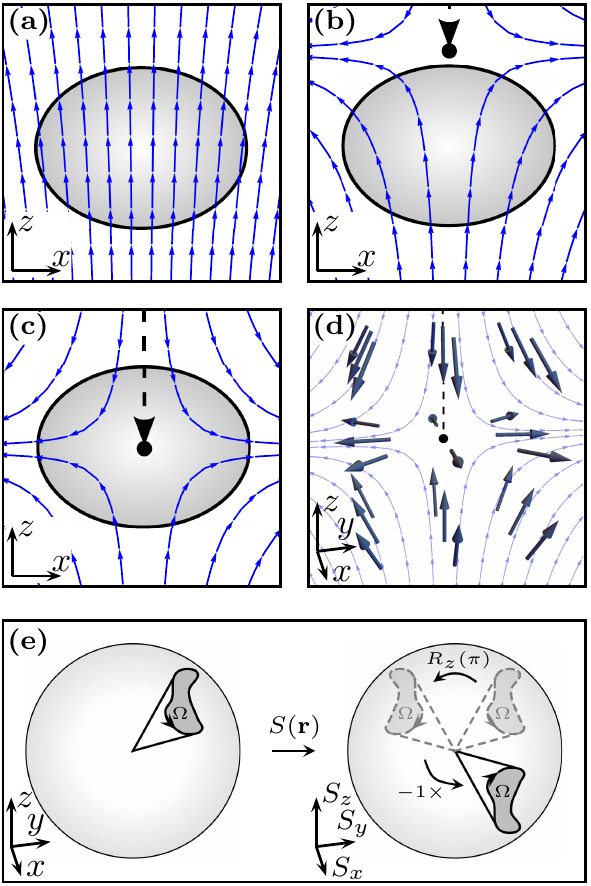}
\caption{(a--c) Control sequence of the external quadrupole magnetic field (blue thin arrows) during the creation process of an isolated monopole (Dirac monopole) in the polar (ferromagnetic) phase of the spin-1 BEC and (d) the theoretical spin configuration generating a Dirac monopole. The zero point of the quadrupole magnetic field, indicated by the black dot, is (a) well above, (b) approaching, and (c) in the middle of the condensate (shaded ellipse). The dashed arrow in (b--d) shows the path traced by the zero point as it is adiabatically brought into the condensate. In (d), the directions of the thick arrows indicate the direction of local spin, aligned with the quadrupole field, at selected points in space. (e) Mapping of a closed path on a spherical surface enclosing a solid angle $\Omega$ from the real space to the spin space in Dirac monopole spin configuration (see Sec.~\ref{sec:berry}). In the spin space, we illustrate how an operation  $-R_z(\pi)$ can be used to implement the mapping.}\label{fig:creation}
\end{figure}

In addition to displaying the scalar properties of a superfluid, spinor BECs exhibit intriguing magnetic order in their spin degrees of freedom. Consequently, the spectrum of topological defects available in spinor systems is rich, including vortices~\cite{Matthews:1999,Leanhardt:2002,Mizushima:2002,Leanhardt:2003,Fetter:2009,Telles:2016}, half-quantum vortices~\cite{Seo:2015}, solitonic vortices~\cite{Donadello:2014}, skyrmions~\cite{Khawaja:2001,Leslie:2009,Choi:2012}, monopoles~\cite{Stoof:2001,Savage:2003,Pietila:2009_2,Pietila:2009,Ray:2014,Ray:2015}, and knots~\cite{Kawaguchi:2008,Hall:2016}. Although many of these topologically stable defects remain essentially intact for the duration of the creation process~\cite{Stoof:2001,Savage:2003,Shin:2004,Pietila:2009_2,Pietila:2009,Ray:2014,Ray:2015,Kawaguchi:2008,Hall:2016}, there may be decay channels rendering them dynamically unstable~\cite{Pu:2001,Ruostekoski:2003,Pietila:2007}.

The natural magnetic phase of an atomic spin-1 BEC is either ferromagnetic or polar~\cite{Ho:1998}, depending on the atomic species. With higher-dimensional internal structure, even more magnetic phases are available~\cite{Ueda:2002}. Interestingly, the spin degree of freedom can give rise to synthetic electromagnetism, in which a part of the BEC order parameter acts as a charged quantum particle in the presence of synthetic electromagnetic potentials arising from spatiotemporal variations of its spinor~\cite{Kawaguchi:2012}. These variations have been introduced through interactions with external laser~\cite{Lin_Nature:2009,Dalibard:2011} and magnetic fields~\cite{Ray:2014}.

We are especially interested in gauge potentials that give rise to a monopole in a synthetic magnetic field. Creation of such a Dirac monopole in a ferromagnetic BEC was recently proposed~\cite{Pietila:2009} and implemented experimentally~\cite{Ray:2014}, providing the first known realization of Dirac's celebrated theory of a charged quantum particle interacting with a fixed magnetic monopole~\cite{Dirac:1931}. We stress that such a Dirac monopole is not accompanied by a topological point defect in the ferromagnetic order parameter.

Another recent experiment~\cite{Ray:2015} reported the first observation of topological point defects, i.e., isolated monopoles, in the polar phase of a spin-1 $^{87}$Rb BEC. The creation process is carried out using an external magnetic field as illustrated in Fig.~\ref{fig:creation}(a--c). Although the point defect is created in the polar phase, the magnetic phase for the ground state of a BEC in the presence of a strong enough external magnetic field is ferromagnetic. It is therefore natural that the condensate eventually decays from the polar phase into a configuration including the ferromagnetic phase. Interestingly, recent theoretical results~\cite{Tiurev:2016} show that a polar BEC with an isolated monopole will evolve into an essentially ferromagnetic spin configuration with a Dirac monopole in its associated synthetic magnetic field. However, experimental studies of this intriguing phenomenon are lacking to date.

We report here the experimental observation of the decay dynamics of an isolated monopole in a $^{87}$Rb spin-1 BEC. In good agreement with theoretical predictions~\cite{Tiurev:2016}, the condensate evolves from the polar to the ferromagnetic phase, giving rise to the decay of the isolated monopole defect in the order parameter into a Dirac monopole in the accompanying synthetic magnetic field. We draw evidence for our conclusions from observations of the column particle densities in different spin states during the decay. The Dirac monopoles are connected to the condensate boundary by nodal lines of vanishing spin density. In previous experiments~\cite{Ray:2014}, these nodal lines were deterministically created as doubly-quantized vortices terminating at the monopole. In contrast, the nodal lines here appear spontaneously as a pair of connected, singly-quantized vortex lines.

We further characterize the Dirac monopole by projecting the condensate spin texture along three perpendicular quantization axes. These projections provide additional evidence that the ground-state spin configuration contains a Dirac monopole in its synthetic magnetic field. We compare our experimental results to numerical simulations, and find them to be in good agreement. Finally, we study the decay rate utilizing the time-dependent magnetization of the condensate.

This paper is organized as follows: In Sec.~\ref{sec:theory}, we provide the necessary theoretical background for describing the monopole defects and their creation protocol in spin-1 BECs. In Sec.~\ref{sec:methods}, we present the employed experimental and numerical methods. Section~\ref{sec:results} is devoted to the experimental results on the decay dynamics and comparison with corresponding numerical simulations. We discuss our results in Sec.~\ref{sec:conclusions}.

\section{\label{sec:theory}Theory}

\subsection{Mean-field theory}

We base our analysis on the zero-temperature mean-field order parameter of the spin-1 BEC
\begin{equation}
\Psi({\bf r},t) = \sqrt{n({\bf r},t)}e^{i\phi({\bf r},t)}\zeta({\bf r},t),\label{eq:op}
\end{equation}
where $n$ is the particle density, $\phi$ is the scalar phase, and $\zeta=\begin{pmatrix}\zeta_{+1},&\zeta_0,&\zeta_{-1}\end{pmatrix}_{\text{Z}}^T$ is a three-component complex-valued spinor satisfying the normalization condition $\zeta^\dagger\zeta=1$. Here, $\zeta$ is expressed in the basis of the $z$-quantized spin states $\left\{ \left| 1 \right>, \left| 0 \right>, \left| -1 \right> \right\}$. In general, the spinors for the ferromagnetic and polar phases in this basis read~\cite{Ho:1998}
\begin{align}
&\zeta_{\text{F}} = \mathcal{U}(\alpha,\beta,\gamma)\begin{pmatrix}1\\0\\0\end{pmatrix}_{\!\text{Z}}= e^{-i\gamma}\begin{pmatrix}
e^{-i\alpha}\cos^2\frac{\beta}{2} \\
\sqrt{2}\cos\frac{\beta}{2}\sin\frac{\beta}{2} \\
e^{i\alpha}\sin^2\frac{\beta}{2}
\end{pmatrix}_{\!\text{Z}}\!,\\
&\text{and},\nonumber\\
&\zeta_{\text{P}}= \mathcal{U}(\alpha,\beta,\gamma)\begin{pmatrix}0\\1\\0\end{pmatrix}_{\!\text{Z}}= \frac{1}{\sqrt{2}}\begin{pmatrix}
-e^{-i\alpha}\sin\beta \\
\sqrt{2}\cos\beta \\
e^{i\alpha}\sin\beta
\end{pmatrix}_{\!\text{Z}}\!,\label{eq:polarspinor}
\end{align}
where $\mathcal{U}=e^{-iF_z\alpha}e^{-iF_y\beta}e^{-iF_z\gamma}$ is the general spin rotation operator with Euler angles $\alpha$, $\beta$, and $\gamma$, and $F_y$ and $F_z$ are the standard dimensionless spin-1 matrices. The subscripts F and P are used to refer to the ferromagnetic and polar phase, respectively.

\subsection{Topological defects in spin-1 Bose--Einstein condensates}

The spinor in the polar phase in Eq.~(\ref{eq:polarspinor}) can alternatively be written as 
\begin{equation}
\zeta_{\text{P}} = \frac{1}{\sqrt{2}}\begin{pmatrix}-d_x+i d_y\\ \sqrt{2}d_z\\d_x + i d_y\end{pmatrix}_{\!\text{Z}}\!,\label{eq:polar}
\end{equation}
where $\hat{\bf d}=(d_x,d_y,d_z)^T=(\cos\alpha\sin\beta,\sin\alpha\sin\beta,\cos\beta)^T$ is a real-valued unit vector defining the direction of nematic order in the condensate, i.e., the polar spinor is in the eigenstate of ${\bf F}\cdot \hat{\bf d}$ with eigenvalue $m_F=0$. Here, ${\bf F}=(F_x,F_y,F_z)^T$ is a vector of dimensionless spin-1 matrices. The polar order parameter may thus be expressed as $\Psi=\sqrt{n}e^{i\phi}\hat{\bf d}$ in the Cartesian basis. We use the vector field $\hat{\bf d}({\bf r},t)$ to characterize magnetic-order-related topological defects in the polar phase.

In the pure ferromagnetic phase, the local spin magnitude is unity, whereas it vanishes in the pure polar phase. Hence, in the ferromagnetic phase, the local average spin 
\begin{equation}
{\bf S}({\bf r}) = \zeta({\bf r})^\dagger{\bf F}\zeta({\bf r}),\label{eq:spin}
\end{equation} 
plays an important role in the formation of possible topological defects.

The order parameter space in the polar phase is given by $\mathcal{O}_{\text{P}}\cong\left[\text{U}(1)\times S^2\right]/\mathbb{Z}_2$~\cite{Zhou:2001}, allowing the existence of singular point defects due to the nontriviality of the second homotopy group, $\pi_2(\mathcal{O}_{\text{P}})\cong\mathbb{Z}$~\cite{Nakahara:2003}. For the ferromagnetic spinor, the order parameter space is $\mathcal{O}_{\text{F}}\cong\text{SO}(3)$~\cite{Kawaguchi:2012}, and hence the second homotopy group is trivial, $\pi_2(\mathcal{O}_{\text{F}})\cong0$, forbidding the existence of topologically stable point defects~\cite{Makela:2003}. As pointed out above, the Dirac monopoles we consider here are not point defects in the ferromagnetic order parameter but in the associated synthetic magnetic field, and may therefore exist at the termination points of vortex lines in the scalar degree of freedom~\cite{Pietila:2009}. Dirac refers to such vortex lines as nodal lines in his original work~\cite{Dirac:1931} since the probability density of the associated wavefunction vanishes at the vortex core.

\subsection{Berry phase and synthetic electromagnetism}\label{sec:berry}
The concept of Berry phase plays an important role in spinor gases. As we show below, it shares a relation with synthetic electromagnetism, which we formally review in Appendix~\ref{app:synth}. The Berry phase may appear in the condensate order parameter due to adiabatic spin rotations or spatially dependent spin configurations. As one traverses around a closed path $\mathcal{C}$ in real space, the accumulated Berry phase due to the spatial dependence of the spinor is given by~\cite{Berry:1984}
\begin{align}
\Theta_{\text{B}}&=i\oint_\mathcal{C} \zeta({\bf r},t)^\dagger\nabla\zeta({\bf r},t)\cdot \text{d}{\bf r}\nonumber\\
&=i\int_\mathcal{S}\nabla\times\left[\zeta({\bf r},t)^\dagger\nabla\zeta({\bf r},t)\right]\cdot \text{d}{\bf S}\nonumber\\
&=q^*\!\int_\mathcal{S}\nabla\times {\bf A}^*({\bf r},t)\cdot \text{d}{\bf S}=\frac{q^*\!}{\hbar}\int_\mathcal{S}{\bf B}^*({\bf r},t)\cdot \text{d}{\bf S},\label{eq:synthflux}
\end{align}
where $\mathcal{S}$ is the area enclosed by $\mathcal{C}$ and the second identity follows from Stokes' theorem. Here, we have introduced the synthetic charge $q^*$, the synthetic vector potential ${\bf A}^*$, and the synthetic magnetic field ${\bf B}^*$, which are discussed in more detail in Appendix~\ref{app:synth}. With these definitions, the accumulated Berry phase equals the Aharonov--Bohm phase that a charged scalar particle with charge $q^*$ would accumulate when moving along $\mathcal{C}$ in a natural magnetic field coinciding with ${\bf B}^*$. Thus the scalar part of the order parameter simulates the behavior of the charged particle.

Let us evalute the Berry phase associated with the Dirac monopole spin configuration. Figure~\ref{fig:creation}(e) shows the mapping from the real space to the spin space in this configuration. A closed path, enclosing a solid angle $\Omega$ on the surface of a sphere in the real space, maps to a path in the spin space enclosing an equal-magnitude solid angle $\Omega$ but traversing in the opposite direction. In the spin-1 case in general, the accumulated Berry phase equals the solid angle in the spin space, when traversed in the negative direction~\cite{Berry:1984,Leanhardt:2002}. Thus we obtain $\Theta_{\mathrm{B}}=\Omega$ for a path traversed in the positive direction in real space. On the other hand, using Eq.~(\ref{eq:synthflux}), we can write $\Theta_{\mathrm{B}}=q^*\!\int_\Omega {\bf B}^*\cdot \text{d}{\bf S}/\hbar=q^*\Phi_{\mathrm{B}}^*/\hbar$, where $\Phi_{\mathrm{B}}^*$ is the synthetic magnetic flux through $\Omega$. Hence, in the Dirac monopole spin configuration, the Berry phase gives rise to the synthetic magnetic flux of a monopole, i.e., $\Phi^*_{\mathrm{B}}=\hbar\Theta_{\mathrm{B}}/q^*=\hbar\Omega/q^*$, independent of the other details of the path than the solid angle.

\subsection{Monopole creation}

We create isolated monopole defects in the polar BEC by applying spin rotations to the spinor $\zeta$ using an external quadrupole magnetic field. The magnetic field is well approximated by 
\begin{equation}
{\bf B}({\bf r}',t)=b_{\text{q}}\left( x'\hat{\bf x}' + y'\hat{\bf y}' - z'\hat{\bf z}' \right) + {\bf B}_{\text{bias}}(t),\label{eq:quadrupole}
\end{equation}
where $b_{\text{q}}$ is the quadrupole field strength, ${\bf B}_{\text{bias}}(t)=(B_{\text{bias},x}(t),B_{\text{bias},y}(t),B_{\text{bias},z}(t))^T$ is a uniform bias field, and the primed Cartesian basis vectors are given by $\left\{\hat{\bf x}', \hat{\bf y}', \hat{\bf z}'\right\}=\left\{\hat{\bf x}, \hat{\bf y}, 2\hat{\bf z}\right\}$. The monopole is created by slowly bringing the zero point of the magnetic field, ${\bf r}'_0(t) = (-B_{\text{bias},x}(t),-B_{\text{bias},y}(t),B_{\text{bias},z}(t))^T/b_{\text{q}}$, into the middle of the condensate [see Fig.~\ref{fig:creation}(a--c)], such that $\hat{\bf d}$ follows the direction of the local magnetic field adiabatically.

The isolated monopole created in this way is a topological point defect defined by the nematic vector $\hat{\bf d}_{\text{m}}=\left( x'\hat{\bf x}' + y'\hat{\bf y}' - z'\hat{\bf z}' \right)/r'$, where $r'=\sqrt{x'^2+y'^2+z'^2}$. It is related to the hedgehog monopole $\hat{\bf d}_{\text{h}}$, in which the nematic vector field points radially outwards from the location of the monopole, through a sign change and $\pi$ rotation about the $z$ axis, $\hat{\bf d}_{\text{h}}=-e^{-iF_z\pi}\hat{\bf d}_{\text{m}}=\left( x'\hat{\bf x}' + y'\hat{\bf y}' + z'\hat{\bf z}' \right)/r'$. Thus, the two configurations, $\hat{\bf d}_{\text{m}}$ and $\hat{\bf d}_{\text{h}}$, both describe a point defect.

The spin configuration hosting a Dirac monopole is defined by the spin vector ${\bf S}_{\text{D}}=\left| {\bf S}_{\text{D}} \right|\left( x'\hat{\bf x}' + y'\hat{\bf y}' - z'\hat{\bf z}' \right)/r'$. Since point defects are not topologically allowed in the ferromagnetic phase of spin-1 BECs, this spin texture is naturally accompanied by one or more nodal lines connecting the boundary of the condensate with the core of the monopole. The nodal lines appear not in the condensate spin texture but in the density. In the previously realized Dirac monopoles~\cite{Ray:2014}, the nodal line is a doubly-quantized vortex connecting to the defect. This doubly-quantized vortex dynamically splits into the more energetically favorable configuration containing two singly-quantized vortices~\cite{Mottonen:2003,Ray:2014}. Indeed, in the ground-state Dirac monopole configuration~\cite{Ruokokoski:2011}, there are two single-quantum vortices connecting to the monopole. Furthermore, these nodal lines have vanishing particle density only in the pure ferromagnetic phase. For the Dirac monopoles arising as a result of the decay of the isolated monopole, the nodal lines are expected to appear as two single-quantum vortices in the partially-depleted total particle density and as more distinct depletions in the spin density~\cite{Tiurev:2016}. This indicates that the nodal lines are partly filled with polar-phase atoms, and may therefore be referred to as polar-core vortices~\cite{Ruokokoski:2011}. Recently, it was theoretically shown that in the absence of any external magnetic fields, the isolated monopole decays into a nodal-line-like polar-core vortex of nonvanishing particle density at the vortex core~\cite{Tiurev:2016_2}.

\section{\label{sec:methods}Methods}
\subsection{\label{sec:experiments}Experimental methods}
The experimental protocol employed in the creation of isolated monopoles is essentially identical to that in Ref.~\cite{Ray:2015}. Our experiments begin with optically trapped $^{87}$Rb atoms prepared in the polar internal state $\hat{\bf z}=(0,1,0)^T_{\text{Z}}$. A typical atom number is $N = 2.1\times10^5$ and the thermal cloud is observed to be negligibly small. The optical trapping frequencies are $\omega_r\approx2\pi\times130$~Hz and $\omega_z\approx2\pi\times170$~Hz. The strength of the gradient field is $b_{\text{q}}=4.3$~G/cm and the field zero is moved into the condensate by linearly ramping a bias field, aligned with $\hat{\bf z}$, from 10~mG to zero in 40~ms. The vector field $\hat{\bf d}$ adiabatically follows the external magnetic field during the slow ramp, resulting in the isolated monopole configuration~$\hat{\bf d}_{\text{m}}$.

Once the isolated monopole is created, we hold the zero point of the quadrupole field in the middle of the condensate for $t_{\text{hold}}$. We vary $t_{\text{hold}}$ for different experimental runs, which provides us with information on the condensate dynamics at different stages of the decay. As the polar phase decays into the ferromagnetic phase, the magnetic order becomes defined by the local spin ${\bf S}$ rather than $\hat{\bf d}$.

After holding the monopole in the presence of the quadrupole field for $t_{\text{hold}}$, we apply a projection ramp along a chosen quantization axis. This is implemented by increasing the bias field to a large value along the quantization axis such that $\left|{\bf B}_{\text{bias}}\right|\gg b_{\text{q}}R$, where $R$ is the effective extent of the condensate. The duration of the projection ramp is approximately 50 $\mu$s and the order parameter is expected to remain essentially unchanged, allowing us to project the condensate state to the eigenstates of the Zeeman Hamiltonian along the chosen quantization axis. The quadrupole contribution to the field is then turned off in a few microseconds and the condensate is released from the optical trap. During the subsequent 23.1-ms free expansion of the cloud, the bias field is adiabatically rotated into the $x$ direction. A 3.5-ms pulse of current applied to the quadrupole field coils produces a magnetic gradient that separates the spin states horizontally. Finally, the bias field is adiabatically rotated to point along the $y$ axis and the condensate cloud with spatially separated spin states is simultaneously imaged along $y$ and $z$. A detailed description of the projection ramp and the associated imaging methods are presented in Ref.~\cite{Ray:2014}.

\subsection{\label{sec:numerics}Numerical methods}
We numerically simulate our experiments by solving the dynamics of the mean-field order parameter $\Psi$ according to the Gross--Pitaevskii (GP) equation and by employing the literature values for the constants. The GP equation reads
\begin{align}
i\hbar \frac{\partial}{\partial t} \Psi ({\bf r},t) =& \left\{h({\bf r},t) + n({\bf r},t)\left[ c_0 + c_2 {\bf S}({\bf r},t)\cdot {\bf F} \right] \vphantom{n^2}\right.\nonumber\\
&\left.-i\Gamma n^2({\bf r},t)\right\}\Psi({\bf r},t),\label{eq:gp}
\end{align}
where $\hbar$ is the reduced Planck constant, $c_0=4\pi\hbar^2(a_0+2a_2)/(3m)$ and $c_2=4\pi\hbar^2(a_2-a_0)/(3m)$ are the constants related to density--density and spin--spin interactions~\cite{Ohmi:1998,Ho:1998}, respectively, with the $s$-wave scattering lengths being $a_0=5.387$~nm and $a_2=5.313$~nm~\cite{vanKempen:2002}, and $m=1.443\times10^{-25}$~kg is the mass of a $^{87}$Rb atom. The three-body recombination rate is $\Gamma=\hbar\,\times\,2.9\,\times\,10^{-30}$~cm$^6/$s~\cite{Burt:1997}. The single-particle Hamiltonian $h$ is given by 
\begin{equation}
h({\bf r},t)=-\frac{\hbar^2}{2m}\nabla^2+V_{\text{opt}}({\bf r}) + g_{F}\mu_{\text B}{\bf B}({\bf r},t)\cdot {\bf F},\label{eq:sp}
\end{equation}
where $V_{\text{opt}}$ is the optical trapping potential, $g_{F}=-1/2$ is the Land\'e $g$-factor, and $\mu_{\text{B}}$ is the Bohr magneton. The quadratic Zeeman shift is not included since it does not have a significant effect on the dynamics considered here~\cite{Tiurev:2016}. The optical trap in the vicinity of the condensate is approximated by a harmonic potential $V_{\text{opt}}=m\left[ \omega_r^2\left(x^2+y^2\right) +\omega_z^2z^2 \right]/2$. The external magnetic field ${\bf B}$ assumes the quadrupole 
configuration given by Eq.~(\ref{eq:quadrupole}).

The ground state of the system is found by a relaxation method~\cite{Press:1994} with the bias fields set along positive $z$ such that $\left| {\bf B}_{\text{bias}}\right|\gg b_{\text{q}}Z$, where $Z$ is the effective axial extent of the condensate along $z$. With the experimental parameters, the relaxation leads to an essentially spin-polarized ferromagnetic phase, $\zeta\approx\begin{pmatrix}1,&0,&0\end{pmatrix}_{\text{Z}}^T$. Subsequently, the spinor components are swapped resulting in $\zeta\approx\begin{pmatrix}0,&1,&0\end{pmatrix}_{\text{Z}}^T$. The experimental protocol is then simulated by solving the corresponding temporal evolution from Eq.~(\ref{eq:gp}). Here, we employ the split operator method together with fast Fourier transforms~\cite{Press:1994}. The computations are carried out on a discretized three-dimensional grid of size $200\!\times\!200\!\times\!200$ using graphics processing units.

\section{\label{sec:results}Results}

In Sec.~\ref{sec:result1}, we show the particle density distributions of the condensate spin states for the isolated monopole in the beginning and during the decay process, as well as for the resultant Dirac monopole. The nodal lines attributed to the Dirac monopole are shown and analyzed in Sec.~\ref{sec:result2}. Section~\ref{sec:result3} is devoted to characterization of the Dirac monopole using different projection axes. In Sec.~\ref{sec:result4}, we present our observations of the rate of the monopole decay process.

\begin{figure}[t!]
\centering
\includegraphics[width=0.45\textwidth]{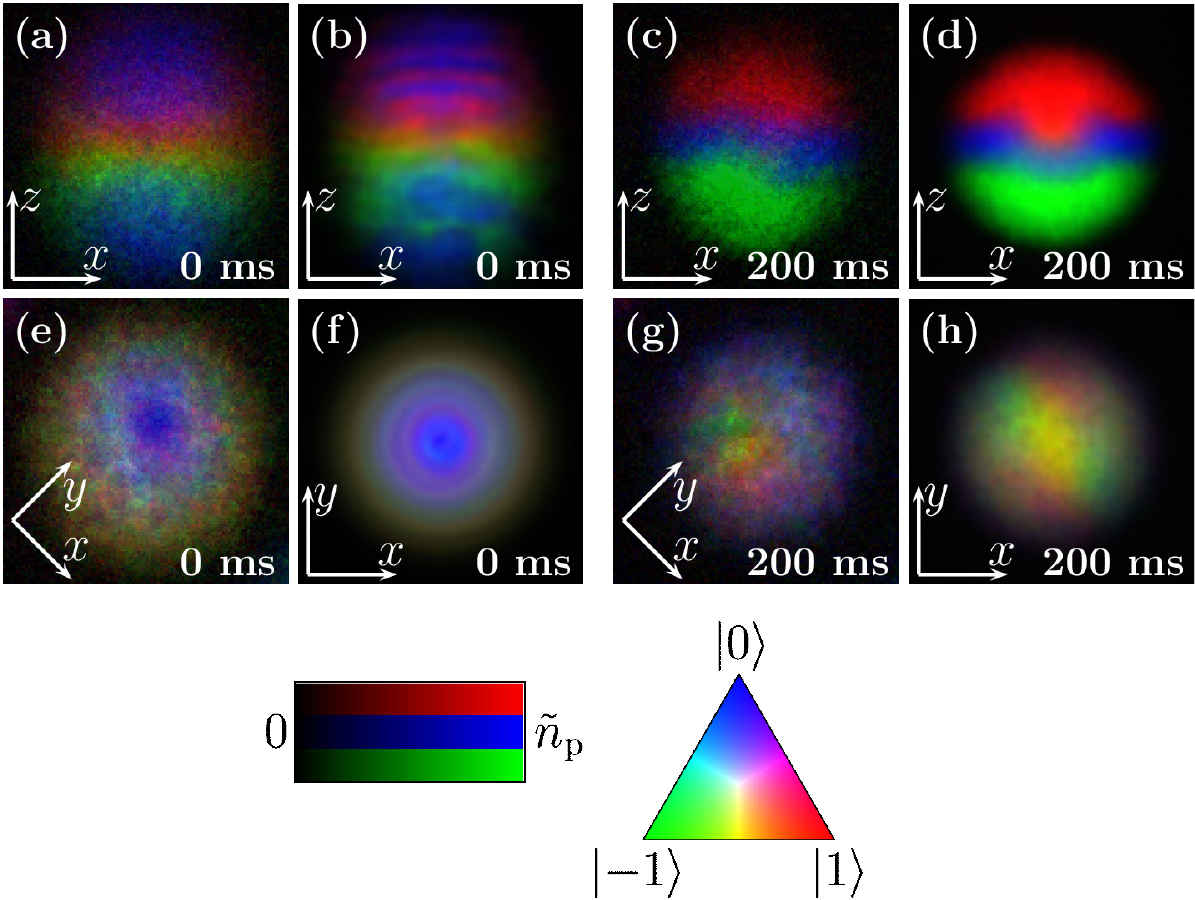}
\caption{Column particle densities in the three $-z$-quantized spin states for (a,b,e,f) the isolated monopole and (c,d,g,h) the dynamically formed Dirac monopole. The experimental data and the corresponding simulation results are shown in panels (a,c,e,g) and (b,d,f,h), respectively. The hold times are indicated in the bottom right corner of each panel. The quadrupole field gradient is $b_{\text{q}}=4.3$~G/cm. The peak particle density is $\tilde{n}_{\mathrm{p}}=8.5\times10^8~\mathrm{cm^{-2}}$ for the images along $y$ in (a--d) and $\tilde{n}_{\text{p}}=1.0\times10^{9}~\text{cm}^{-2}$ for the images along $z$ in (e--h). The field of view in each panel is $228\!\times\!228~\mu\mathrm{m^2}$.}\label{fig:decay}
\end{figure}

\subsection{\label{sec:result1}Decay of an isolated monopole}

\begin{figure}[t!]
\centering
\includegraphics[width=0.45\textwidth]{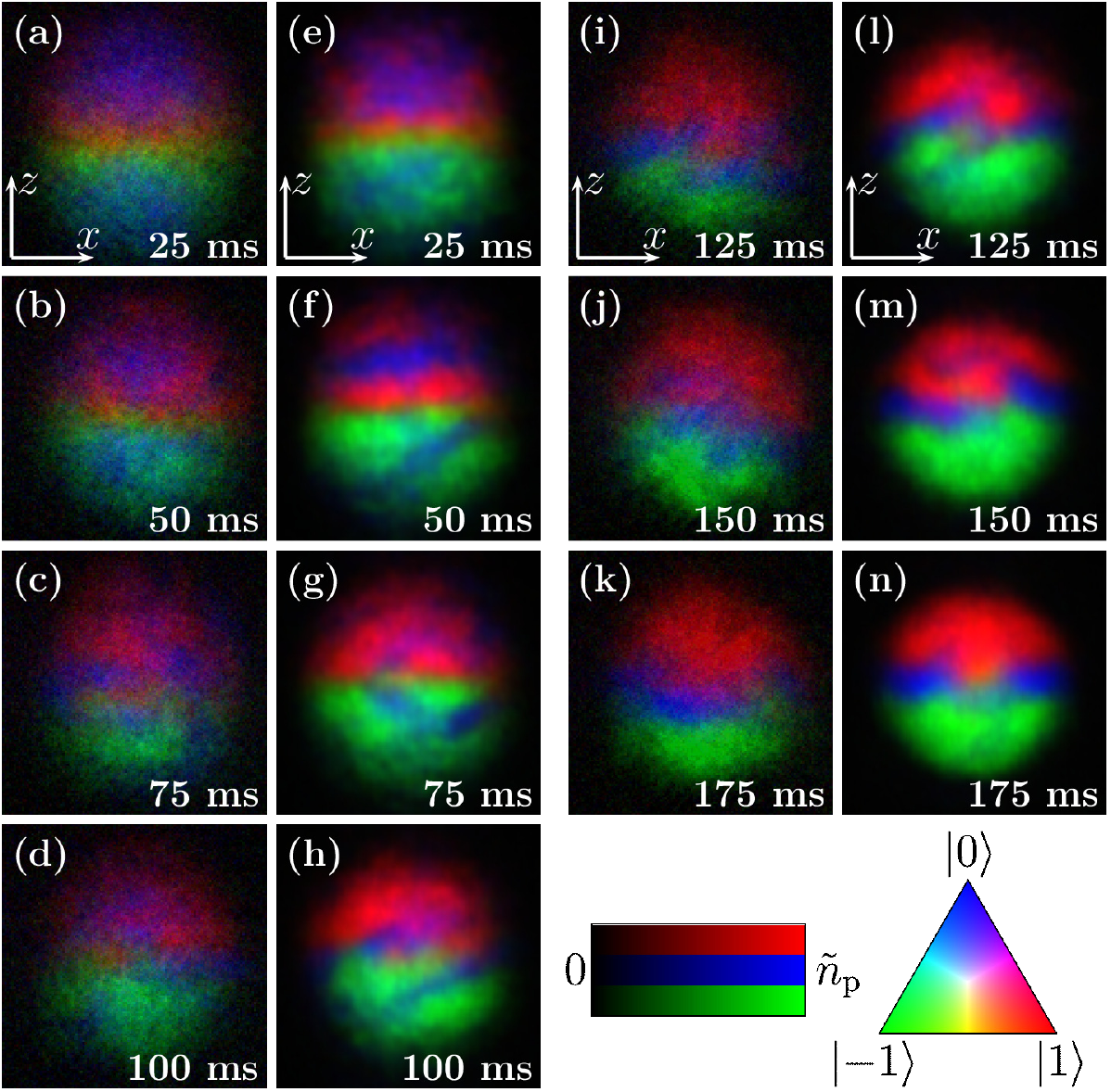}
\caption{Column particle densities imaged along $y$ in the three $-z$-quantized spin states during the decay of an isolated monopole for (a--d,i--k) experiments and (e--h,i--k) simulations. The hold times are indicated in the bottom right corner of each panel. The quadrupole field gradient is $b_{\text{q}}=4.3$~G/cm. The peak particle density is $\tilde{n}_{\mathrm{p}}=8.5\times10^8~\mathrm{cm^{-2}}$ and the field of view in each panel is $228\!\times\!228~\mu\mathrm{m^2}$.}\label{fig:decay_intermediate}
\end{figure}

Figure \ref{fig:decay} shows the column particle densities in the different spin states of the experimentally and numerically created isolated monopoles and those of the Dirac monopoles after the decay process. Figure~\ref{fig:decay_intermediate} shows the particle densities during the decay. Although the agreement between the experiments and the simulations is good in general, the constant loss of atoms from the trap and the anharmonicity of the optical potential is not included in the simulations, resulting in a slight disagreement in the contrast of the images. Hereafter, the images taken along $y$ and $z$ will be referred to as side and top images, respectively. 

At $t_{\text{hold}}=0$~ms [see Fig.~\ref{fig:decay}(a,b,e,f)], the $-z$-quantized spinor component $\zeta_0$ occupies the top and the bottom parts of the condensate cloud, while the spinor components $\zeta_{\pm1}$ occupy the middle region with partial overlap as observed in the side images. The structure is consistent with the analytical column particle densities corresponding to the polar spinor with the vector field $\hat{\bf d}$ oriented along the local quadrupole magnetic field [see Eq.~(\ref{eq:polarspinor})].  In the experiments and simulations, the $\zeta_{\pm1}$ components do not fully overlap due to the fact that the projection ramp is applied before the quadrupole field is completely switched off. 
In the top images, the blue region corresponds to density depletions in the $\zeta_{\pm1}$ components which form a ring around the $\zeta_0$ component. These are the characteristic indications that an isolated monopole has been created in the condensate cloud~\cite{Ray:2015}.

At the early stages of the decay process, at $t_{\text{hold}}=25~\text{ms}$ [see Fig.~\ref{fig:decay_intermediate}(a,e)], the monopole structure is still visible, with the $\zeta_{+1}$ and $\zeta_{-1}$ spinor components being displaced very slightly to the top and bottom parts of the condensate, respectively. At $t_{\text{hold}}=50~\text{ms}$ [see Fig.~\ref{fig:decay_intermediate}(b,f)], the $\zeta_{\pm1}$ components have continued to displace, and in addition, the $\zeta_{0}$ component has moved towards the center region. Here, the condensate is far from the pure polar phase, and consequently the isolated monopole structure is not well defined.

For $125~\text{ms}\le t_{\text{hold}}\le150~\text{ms}$ [see Fig.~\ref{fig:decay_intermediate}(i,j,l,m)], the condensate is almost in the pure ferromagnetic phase, as indicated by the well-separated spin states. Finally, at $t_{\text{hold}}=200~\text{ms}$ [see Fig.~\ref{fig:decay}(c,d,g,h)], the density profile of the spin states accurately corresponds to a Dirac monopole~\cite{Ray:2014}. For this hold time, the $-z$-quantized spinor components $\zeta_{+1}$ (spin down), $\zeta_{0}$ (spin horizontal), and $\zeta_{-1}$ (spin up) occupy the top, middle, and bottom parts of the condensate, respectively. This is consistent with the spin aligned along the quadrupole magnetic field as illustrated in Fig.~\ref{fig:creation}(d). 
Thus, in both the simulations and the experiments, the condensate has continuously decayed from the isolated monopole to the spin configuration expected for a Dirac monopole.

\subsection{\label{sec:result2}Observation of nodal lines}

\begin{figure}[t!]
\centering
\includegraphics[width=0.5\textwidth]{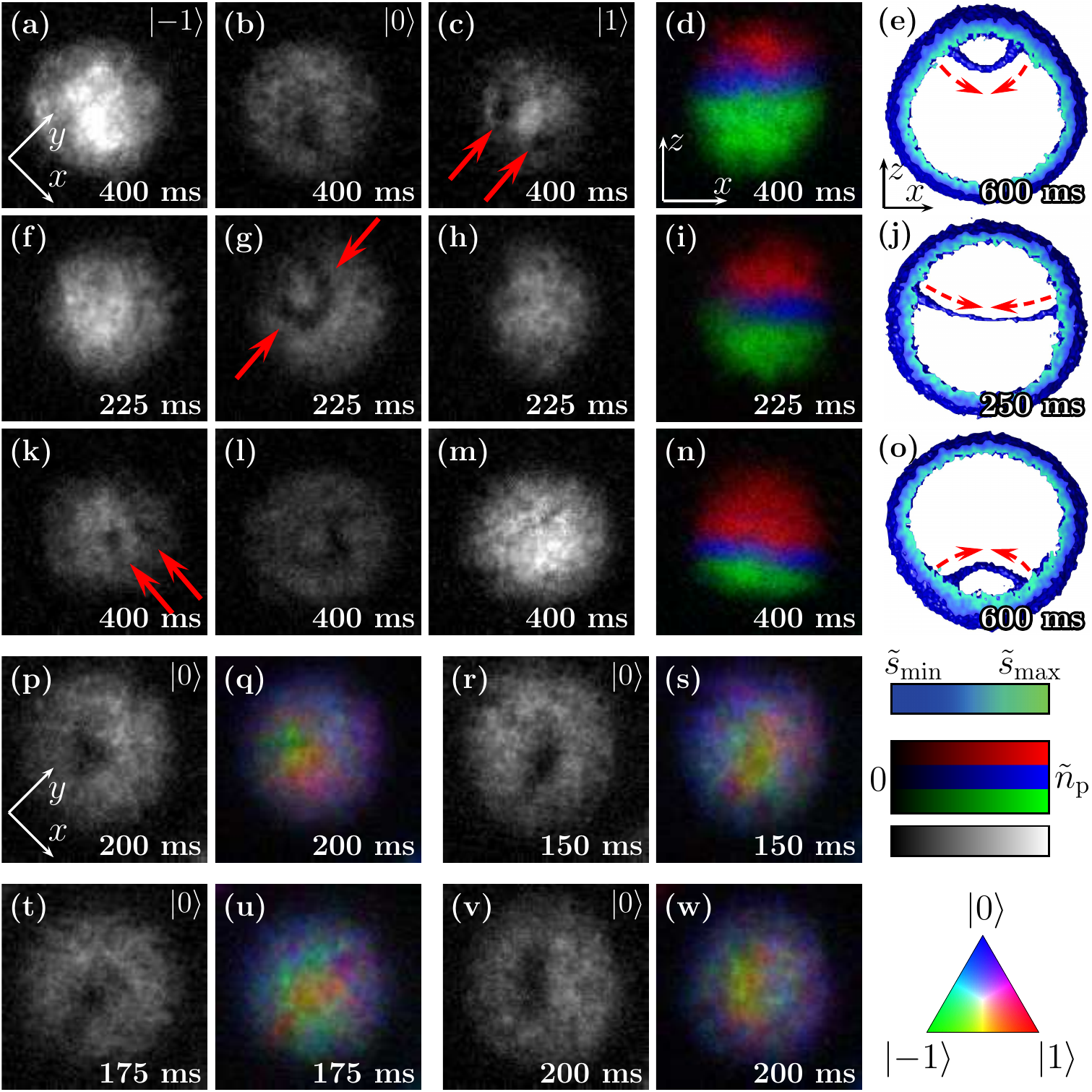}
\caption{Experimental column particle densities of $-z$-quantized spin states showing nodal lines associated with the Dirac monopole. In (a--o), the first, second, and third columns correspond to the top images of $\zeta_{-1}$, $\zeta_{0}$, and $\zeta_{+1}$ spinor components, respectively, the fourth column shows the related composite side images, and the fifth column shows the corresponding numerically obtained spin densities $\tilde{s}=\left|\Psi^\dagger{\bf F}\Psi\right|$ before the projection ramp. In (a--e), the monopole is located in the upper half of the condensate, in (f--j) in the middle, and in (k--o) in the lower half. The solid red arrows in (c,g,k) indicate the locations of the nodal lines and the dashed red arrows in (e,j,o) indicate the direction of the vorticity of the nodal line. In (p--w), each set of two adjacent images is extracted from an individual experiment in which the monopole is located in the middle of the condensate, such that (p,r,t,v) show the column particle density of the $\zeta_0$ spinor component and (q,s,u,w) show the corresponding composite top image, respectively. The data for (p,q) are the same as in Fig.~\ref{fig:decay}(g). The hold times are indicated in the bottom right corner of each panel. 
The fields of view are as in Fig.~\ref{fig:decay}, and the peak particle densities are (a--c,f--h,k--m,p,r,t,v) $\tilde{n}_{\text{p}}=5.0\times10^8~\text{cm}^{-2}$, (d,i,n) $\tilde{n}_{\text{p}}=8.5\times10^8~\text{cm}^{-2}$, and (q,s,u,w) $\tilde{n}_{\text{p}}=1.0\times10^9~\text{cm}^{-2}$. The simulation data in (e,j,o) is shown for a region of $15\times15$~$\mu$m$^2$ with the thickness of 5.8~$\mu$m along $y$. The shown minimum and maximum spin densities are $\tilde{s}_{\text{min}}=2.1\times10^8~\text{cm}^{-3}$ and $\tilde{s}_{\text{max}}=10.1\times10^8~\text{cm}^{-3}$.}\label{fig:nodal}
\end{figure}

The column particle densities of the condensates showing the nodal lines associated with the Dirac monopole are presented in Fig.~\ref{fig:nodal}. As noted above, Dirac's theory requires the existence of nodal lines for a quadrupolar spin texture, as does the topological constraint forbidding the existence of a point defect in the ferromagnetic phase. In agreement, we observe a pair of singly-quantized vortex lines to spontaneously appear during the decay, connecting the core of the monopole to the condensate boundary. The spontaneous emergence of the nodal lines highlights their topological origin. In the experiments detailed in Ref.~\cite{Ray:2014}, the nodal lines were deterministically created as doubly-quantized vortices terminating at the monopole. Here, in contrast, the nodal lines do not emerge as doubly-quantized vortices, because the configuration with two singly-quantized vortices is energetically more favorable~\cite{Ruokokoski:2011}. As confirmed by our simulations, the singly-quantized vortices have opposite circulations, and hence can be considered as a single vortex line reversing its winding number at the monopole. Furthermore, it is energetically favorable for the nodal lines to minimize their length~\cite{Ruokokoski:2011} such that they tend to terminate at the condensate boundary that is closest to the monopole. 

The nodal lines are observed in roughly 90\% of the one hundred different experiments conducted for $t_{\text{hold}}\ge150~\text{ms}$. We attribute the missed nodal line observations in some experimental realizations to occasional magnetic field excursions that displace the monopole to positions very close to the radial edge, where the spin configuration is appropriate to the monopole but the nodal lines are too short to resolve.

As shown in Fig.~\ref{fig:nodal}(a--e), if the monopole is located in the upper half of the condensate, the nodal line is typically observed in the top image as two density-depleted holes in the $-z$-quantized $\zeta_{+1}$ spinor component together with a depleted region in the $\zeta_{0}$ component. These density depletions correspond, in fact, to a nodal line extending through the two spatially separated spin states, as is evident from the corresponding numerically obtained spin density in Fig.~\ref{fig:nodal}(e). The monopole is always located in the $\zeta_{0}$ component and hence this component must have a density-depleted region. Importantly, if the monopole is located in the lower half of the condensate, as shown in Fig.~\ref{fig:nodal}(k--o), the nodal line extends to the bottom of the condensate and is observed as density depletions in the $\zeta_{-1}$ and $\zeta_{0}$ components. If the monopole is created in the middle of the condensate, the nodal lines are typically oriented in the $xy$ plane, and they are observed as density-depleted lines in the $\zeta_0$ component, as shown in Fig.~\ref{fig:nodal}(f--j,p--w). In many experimental realizations, the horizontal nodal lines partially extend into the $\zeta_{+1}$ or $\zeta_{-1}$ components. For this reason, the typical observation of the horizontal nodal line is not as distinct as in Fig.~\ref{fig:nodal}(g). 
These observations are in agreement with theoretical expectations~\cite{Ruokokoski:2011,Tiurev:2016} for nodal lines associated with the Dirac monopole. 

\begin{figure}[t!]
\centering
\includegraphics[width=0.45\textwidth]{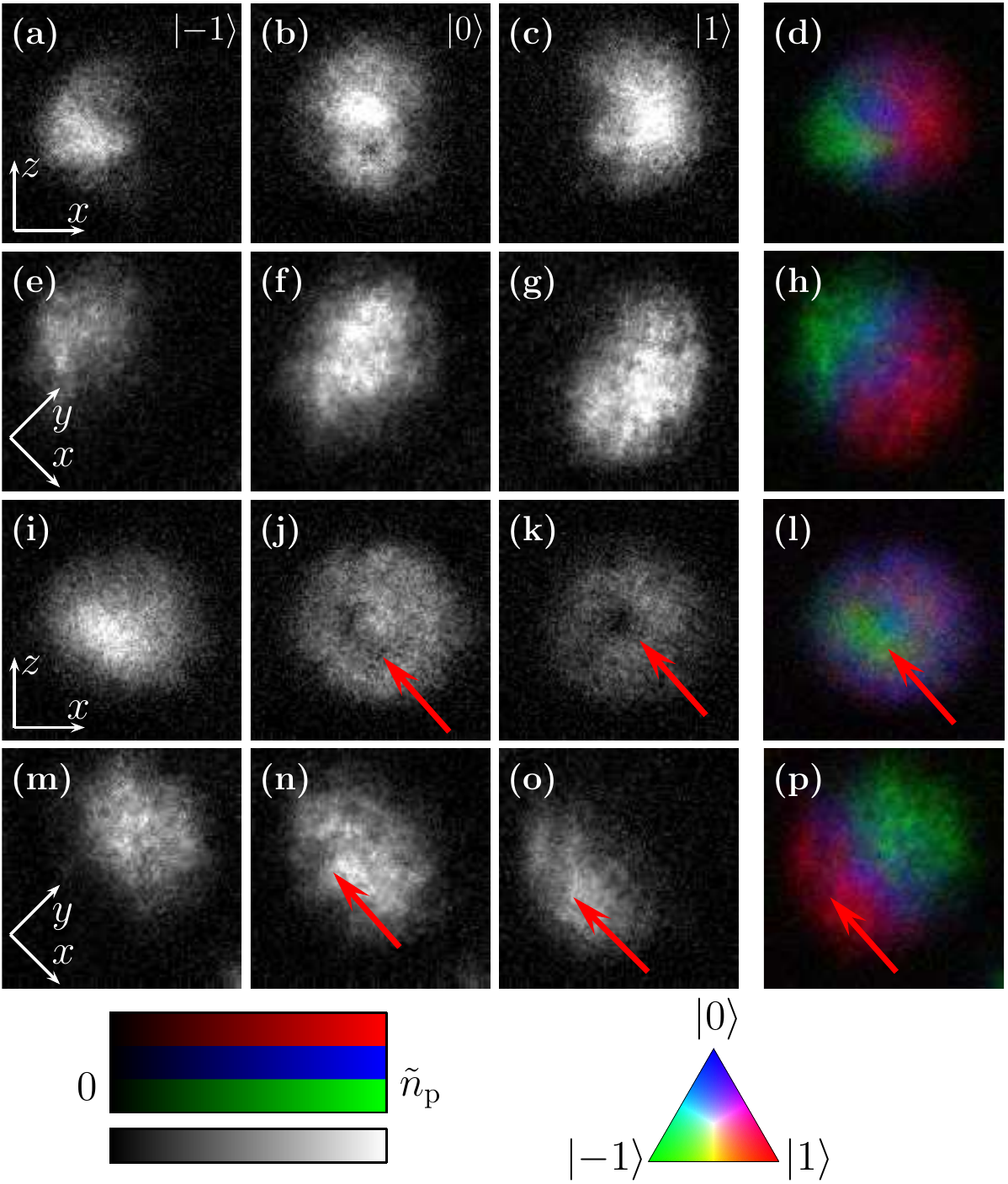}
\caption{Experimental column particle densities of (a--h) $x$- and (i--p) $-y$-quantized spin states. First, second, and third columns correspond to spinor components $\zeta_{-1}$, $\zeta_{0}$, and $\zeta_{+1}$, respectively, and the fourth column shows the corresponding composite image. Here, the hold time is $t_{\text{hold}}=200~\text{ms}$. The red arrows indicate the locations of the nodal lines. The fields of view are as in Fig.~\ref{fig:decay}, and the peak particle densities are (a--c,e--g,i--k,m--o) $\tilde{n}_{\text{p}}=5.0\times10^8~\text{cm}^{-2}$, (d,l) $\tilde{n}_{\text{p}}=8.5\times10^8~\text{cm}^{-2}$, and (h,p) $\tilde{n}_{\text{p}}=1.0\times10^9~\text{cm}^{-2}$.}\label{fig:xyprojs}
\end{figure}

Additional vortices can sometimes appear in the condensate during the BEC creation process or due to the presence of oscillating magnetic fields during the decay (see Appendix~\ref{app:vortex}). These accidental vortices appear in all three spinor components and align themselves with the $z$ axis to minimize their length. In contrast, the nodal lines we observe extend from the monopole core to the nearest boundary. 
Our ability to deterministically control the location of the nodal lines by positioning the magnetic field zero is a strong indication that the nodal lines are not accidental vortices.

\subsection{\label{sec:result3}Characterization of the spin configuration of a Dirac monopole}

The particle densities of the $x$- and $-y$-quantized spin states for $t_{\text{hold}}=200~\text{ms}$ are shown in Fig.~\ref{fig:xyprojs}. We observe that the spin configuration corresponds to that of the Dirac monopole independent of the chosen quantization axis. These observations extend those of Ref.~\cite{Ray:2014}, in which the spin configuration associated with the Dirac monopole was only characterized with the quantization axis parallel to $z$. Furthermore, for the specific $-y$ projection shown in Fig.~\ref{fig:xyprojs}(i--p), the nodal line extends from the origin toward the $-y$ axis and manifests itself as a density depletion in the $\zeta_0$ and $\zeta_{+1}$ spinor components. 

\subsection{\label{sec:result4}Rate of the monopole decay}

We investigate the time scales related to the decay process by calculating a magnetization parameter along $z$, $M_z$, for different hold times $t_{\text{hold}}$. We define the magnetization parameter 
as 
\begin{equation}
M_z=\frac{1}{N}\int\!\text{d}z\int\!\text{d}x \left| n^{y}_{+1}(x,z) - n^{y}_{-1}(x,z)\right|,
\end{equation}
where $n^{y}_{m_z}=\int\!\text{d}y\,n\left| \zeta_{m_z}\right|^2$ is the $y$-integrated particle density in the spin state corresponding to the magnetic quantum number $m_z$. The $-z$-quantized $\zeta_{\pm1}$ spinor components move in opposite vertical directions due to the nonzero gradient present during the projection ramp. Thus, we apply a compensating shift in the vertical direction such that for $t_{\text{hold}}=0~\text{ms}$ the $\zeta_{\pm1}$ components overlap, yielding a minimum value for $M_z$. An identical correction is applied to all the data. As the condensate decays into the ferromagnetic phase, the magnetization parameter increases. 

\begin{figure}[t]
\centering
\includegraphics[width=0.475\textwidth]{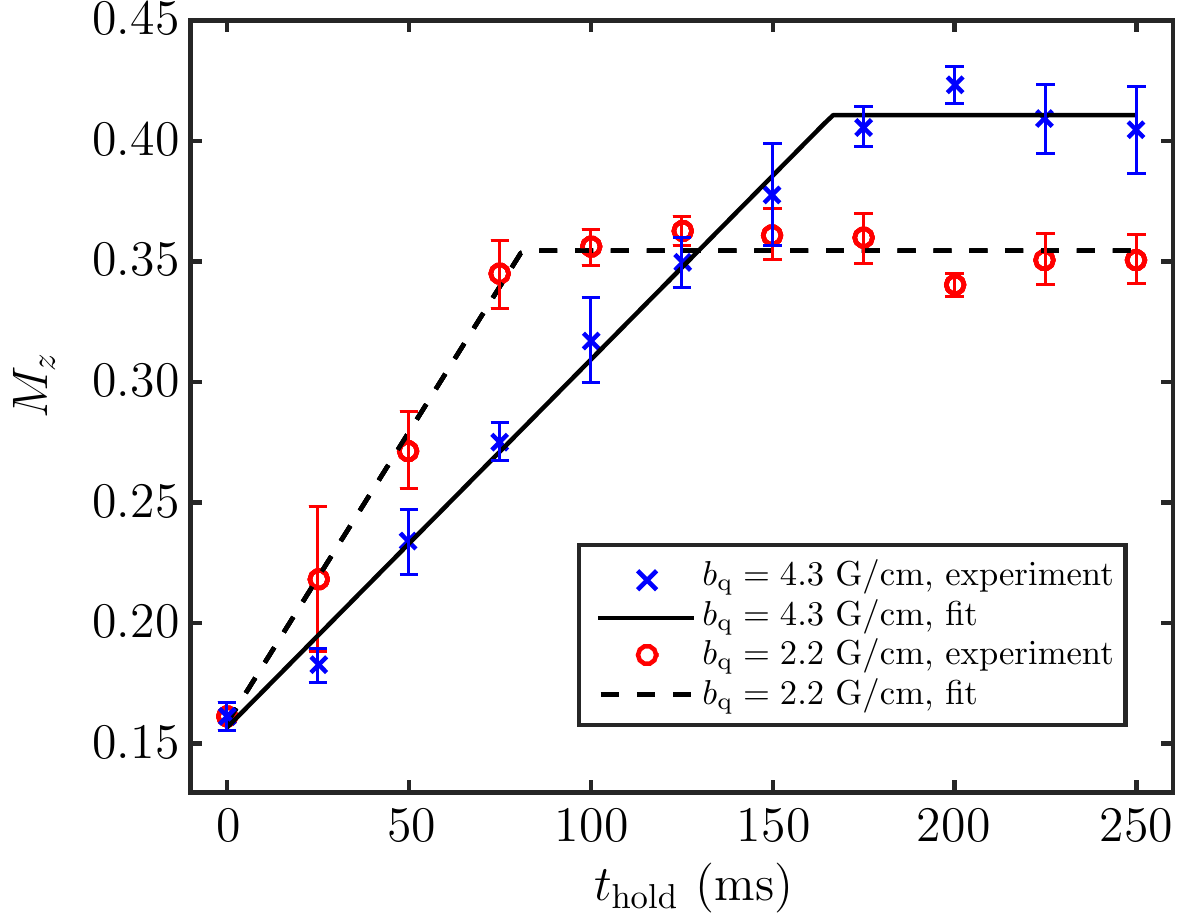}
\caption{Magnetization parameter $M_z$ as a function of hold time $t_{\text{hold}}$. Blue crosses and red circles represent the mean value of ten individual experiments with magnetic field gradients $b_\text{q}=4.3$~G/cm and $b_\text{q}=2.2$~G/cm, respectively. The error bars indicate the uncertainty of two standard deviations of the means. Solid and dashed lines are piecewise linear fitting functions with constant ends.}\label{fig:magnet}
\end{figure}

Figure~\ref{fig:magnet} shows the magnetization parameter of the BEC as a function of the hold time in the presence of two different gradient field strengths, 2.2~G/cm and 4.3~G/cm. In both cases the isolated monopole is created with $b_\text{q}=4.3~\text{G/cm}$, after which $b_\text{q}$ is linearly ramped to its chosen value during the first $10~\text{ms}$ of the evolution. During the ramping of $b_\text{q}$, we adjust the bias fields accordingly to keep the field zero approximately centered in the middle of the condensate. With the weaker gradient, $M_z$ reaches its asymptotic value of approximately 0.35 at $t_{\text{hold}}\approx80~\text{ms}$ with an approximate rate 2.4~$1/\text{s}$. For $t_{\text{hold}}\ge100$~ms we can identify the Dirac monopole spin configuration and the nodal lines from the experimental particle densities (data not shown). With the stronger gradient, the asymptotic value of approximately 0.41 is reached at $t_{\text{hold}}\approx170~\text{ms}$ with an approximate rate 1.5~$1/\text{s}$. Assuming an ideal quadrupole spin configuration, as well as a Thomas--Fermi distribution with the ratio of the radii being $Z/R\approx 1.2$, we obtain an estimate for the theoretical maximum value $M_z\approx0.55$. The saturation values of $M_z$ should be compared to this maximum value.

Our observations indicate that the decay process is slower in the presence of a stronger field gradient. This is attributed to the decreased spatial overlap between the emerging spin domains with stronger gradient~\cite{Tiurev:2016}. Differences in the degree of spatial overlap of the spin domains is also the reason for differences in the asymptotic values of the magnetization parameter for different magnetic field gradients.

\section{\label{sec:conclusions}Conclusion}

We have experimentally studied the decay dynamics of isolated monopoles in spin-1 BECs of $^{87}$Rb atoms. While the condensate evolves from the polar to the ferromagnetic phase, the isolated monopole decays into a spin configuration with a Dirac monopole in its synthetic magnetic field. These results are obtained by analyzing the experimental column particle densities of all spin states projected along different axes. We have also identified spontaneously-appearing nodal lines associated with the emergent Dirac monopole. Numerical simulations are in agreement with these experimental results with no free parameters. The decay of the monopole is observed to be faster in weaker magnetic fields. 

To date, the experimental studies of spatially-localized monopoles in the context of spinor BECs have been limited to two publications~\cite{Ray:2014,Ray:2015} (see also Ref.~\cite{Sugawa:2016}). Our work further verifies the existence of both isolated and Dirac monopoles through reproduction and in-depth characterization. For future experiments, even more precise control of the experimental parameters is desirable in order to probe the delicate structure of other types of topological defects appearing in spinor BECs. Future studies on the stability and the dynamics of, e.g., knot solitons~\cite{Hall:2016}, could be first steps in this direction. Furthermore, the experimental realization of the vortex pump in spinor BECs remains a future challenge requiring precise control of the experimental parameters that determine the dynamics of the atomic cloud~\cite{Mottonen:2007}.

\begin{acknowledgments}
We acknowledge funding by the National Science Foundation (grant PHY-1519174), by the Academy of Finland through its Centres of Excellence Program (grant No. 251748 and No. 284621) and Grant No. 308071, by the European Research Council under Consolidator Grant No. 681311 (QUESS), by the Magnus Ehrnrooth Foundation, by the Education Network in Condensed Matter and Materials Physics, and by the KAUTE foundation through its Researchers Abroad program. CSC--IT Center for Science Ltd. (Project No. ay2090) and Aalto Science-IT project are acknowledged for computational resources. 
\end{acknowledgments}

\appendix

\section{Synthetic electromagnetism}\label{app:synth}

In our system of charge-neutral alkali atoms, spatiotemporal variations in the spinor give rise to synthetic electromagnetism. This is revealed by writing the Gross--Pitaevskii equation, which describes the dynamics of the mean-field order parameter $\Psi=\psi\zeta$, in a form analogous to the Schr\"odinger equation for a scalar charged particle. In this representation, the scalar part of the order parameter $\psi$ plays the role of the wavefunction of the charged particle.

The synthetic vector and scalar potentials acting on the scalar part of the wavefunction assume the forms~\cite{Kawaguchi:2012}
\begin{equation}\label{eq:synthetic_pot1}
{\bf A}^*({\bf r},t) = i\zeta({\bf r},t)^\dagger\nabla\zeta({\bf r},t)/q^*,
\end{equation}
and
\begin{equation}\label{eq:synthetic_pot2}
\Phi^*({\bf r},t) = -i\zeta({\bf r},t)^\dagger\partial_t\zeta({\bf r},t)/q^*,
\end{equation}
respectively. As described in Sec.~\ref{sec:berry}, the synthetic vector potential can be identified with the local Berry connection, which is not a physically observable quantity. Indeed, the potentials are gauge-dependent: we may choose $\tilde{\psi}=\psi e^{-i\eta}$ and $\tilde{\zeta}=e^{i\eta}\zeta$ for a scalar function $\eta$. The mean-field order parameter, and hence all physical observables, remain unchanged under this transformation. However, in this gauge, the synthetic electromagnetic potentials are written as $\tilde{\bf A}^* ={\bf A}^* -\nabla\eta/q^*$ and $\tilde{\Phi}=\Phi + \partial_t\eta/q^*$. 

The synthetic potentials in turn give rise to synthetic electric and magnetic fields 
\begin{equation}\label{eq:synthetic_em1}
{\bf E}^*({\bf r},t) = -\hbar\left[ \nabla\Phi^*({\bf r},t) + \partial_t{\bf A}^*({\bf r},t) \right],
\end{equation}
and
\begin{equation}\label{eq:synthetic_em2}
{\bf B}^*({\bf r},t) = \hbar\left[ \nabla\times{\bf A}^* ({\bf r},t) \right],
\end{equation}
respectively. The related physically observable quantities are the superfluid velocity and the superfluid vorticity, which are expressed as
\begin{equation}\label{eq:synthetic_em3}
{\bf v}_{\text{s}}({\bf r},t) = \frac{\hbar}{m}\left[ \nabla\phi({\bf r},t) - q^*{\bf A}^*({\bf r},t) \right].
\end{equation}
and 
\begin{equation}\label{eq:synthetic_em4}
{\bf \Omega}_{\text{s}}({\bf r},t) = \nabla\times{\bf v}_{\text{s}}({\bf r},t),
\end{equation}
respectively. The superfluid vorticity ${\bf \Omega}_{\text{s}}$ is in fact identical to ${\bf B}^*$ almost everywhere. The singularities in $\phi$ are carried over to ${\bf \Omega}_{\text{s}}$, whereas ${\bf B}^*$ can be made singularity-free with suitable gauge choices in $\eta$ (see Supplementary Information of Ref.~\cite{Ray:2014}).


\section{Emergence of vortices due to moving magnetic field zero point}\label{app:vortex}

\begin{figure}[h]
\centering
\includegraphics[width=0.45\textwidth]{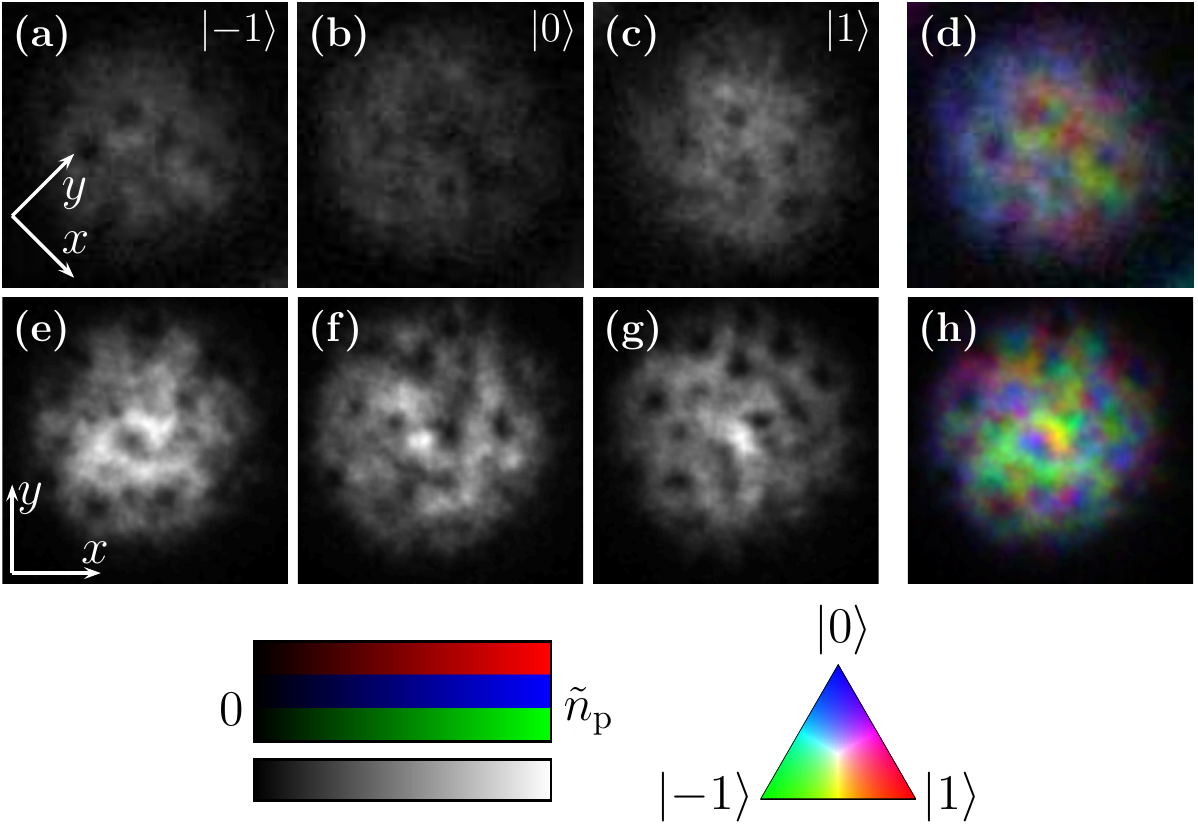}
\caption{(a--d) Experimental column particle densities of $-z$-quantized spin states and (e--h) numerical column particle densities with a circularly rotating field zero in the $xy$ plane corresponding to the magnetic field peak-to-peak amplitude of 1~mG and 60-Hz oscillation frequency. Here, the hold time is $t_{\text{hold}}=100~\text{ms}$. First, second, and third columns correspond to the spinor components $\zeta_{-1}$, $\zeta_{0}$, and $\zeta_{+1}$, respectively, and the fourth column shows the corresponding composite images. In each panel, the peak particle density is $\tilde{n}_{\text{p}}=1.0\times10^{9}~\text{cm}^{-2}$ and the field of view is $228\!\times\!228~\mu\mathrm{m^2}$.}\label{fig:vortices}
\end{figure}

The addition of an oscillating magnetic field at the 60-Hz power line frequency causes the location of the zero point of the magnetic field to oscillate about its central position. It introduces many additional singly-quantized vortices in the condensate during the decay of the isolated monopole, as shown in Fig.~\ref{fig:vortices}. Qualitatively similar results with approximately 15 vortices are obtained in the simulations under conditions in which the magnetic field zero rotates about the center of the condensate in the $xy$ plane. Note that the comparison between theory and experiment in Fig.~\ref{fig:vortices} is qualitative because the experimental trajectory of the field zero was not characterized in detail. In contrast to the nodal line studied in Fig.~\ref{fig:nodal}, which is typically bent and not aligned with the $z$ axis, the additional vortices observed here are consistently aligned with the $z$ axis for a well-centered Dirac monopole and tend to appear in all three spinor components. Similar emergence of vortices has previously been observed by driving the zero point of the external magnetic field outside the condensate, giving rise to an effective Lorentz force~\cite{Choi:2013}.

\bibliography{decay}
\bibliographystyle{prx}

\end{document}